\documentclass[preprint,12pt]{aastex}

\shorttitle{{\it EXIST} GRB Sensitivity}

\begin{document}

\title{{\it EXIST}'s Gamma-Ray Burst Sensitivity}
\author{D. L. Band\altaffilmark{1,2}, J. E. Grindlay\altaffilmark{3},
J.~Hong\altaffilmark{3}, G.~Fishman\altaffilmark{4},
D.~H.~Hartmann\altaffilmark{5}, A.~Garson
III\altaffilmark{6}, H.~Krawczynski\altaffilmark{6},
S.~Barthelmy\altaffilmark{7}, N.~Gehrels\altaffilmark{7},
G.~Skinner\altaffilmark{1,8}}
\altaffiltext{1}{CRESST and Astroparticle Physics
Laboratory, Code 661, NASA/Goddard Space Flight Center,
Greenbelt, MD 20771}
\altaffiltext{2}{Center for Space Sciences and Technology,
University of Maryland, Baltimore County, 1000 Hilltop
Circle, Baltimore, MD 21250}
\altaffiltext{3}{Harvard-Smithsonian Center for
Astrophysics, 60 Garden St., Cambridge, MA}
\altaffiltext{4}{NASA Marshall Space Flight Center, NSSTC,
VP-62, 320 Sparkman Drive, Huntsville, AL 35805}
\altaffiltext{5}{Clemson University, Clemson, SC 29634}
\altaffiltext{6}{Washington University in St. Louis, 1 Brookings
Dr., CB 1105, St. Louis, MO 63130}
\altaffiltext{7}{Astroparticle Physics Laboratory, Code 661,
NASA/Goddard Space Flight Center, Greenbelt, MD 20771}
\altaffiltext{8}{Department of Astronomy, University of
Maryland, College Park, MD 20742}
\email{David.L.Band@nasa.gov}

\begin{abstract}

We use semi-analytic techniques to evaluate the burst sensitivity of
designs for the {\it EXIST} hard X-ray survey mission.  Applying
these techniques to the mission design proposed for the Beyond
Einstein program, we find that with its very large field-of-view and
faint gamma-ray burst detection threshold, {\it EXIST} will detect
and localize approximately two bursts per day, a large fraction of
which may be at high redshift. We estimate that {\it EXIST}'s
maximum sensitivity will be $\sim 4$~times greater than that of {\it
Swift}'s Burst Alert Telescope.  Bursts will be localized to better
than 40~arcsec at threshold, with a burst position as good as a few
arcsec for strong bursts. {\it EXIST}'s combination of three
different detector systems will provide spectra from 3~keV to more
than 10~MeV.  Thus, EXIST will enable a major leap in the
understanding of bursts, their evolution, environment, and utility
as cosmological probes.
\end{abstract}


\section{Introduction}

In its quest to find black holes throughout the universe,
the {\it Energetic X-ray Imaging Survey Telescope (EXIST)}
will detect, localize and study a large number of gamma-ray
bursts, events thought to result from the birth of stellar-mass
black holes.  We present the methods used to calculate {\it
EXIST}'s capabilities as a gamma-ray burst detector; we use the
{\it EXIST} design evaluated by the National Research Council's
`Committee on NASA's Beyond Einstein Program:  An Architecture
for Implementation' (2007; see also Grindlay 2007).  The
combination of large detector area, broad energy coverage,
and wide field-of-view (FOV) will result in the detection
of a substantial number of bursts with a flux distribution
extending to fainter fluxes than that of previous missions.
Thus {\it EXIST} should detect high redshift bursts,
perhaps even bursts resulting from the death of Pop~III
stars. {\it EXIST}'s imaging detectors will localize the
bursts, while the combination of detectors, both imaging
and non-imaging, will result in well-determined spectra
from 3~keV to well over 10~MeV.

In this paper we first describe the {\it EXIST} mission design
(\S 2), emphasizing aspects relevant to burst detection.
Then we present the sensitivity methodology (\S 3), which
we apply to the individual coded mask sub-telescopes (\S
4). {\it EXIST} will consist of arrays of these detectors
with overlapping FOVs, and the overall mission sensitivity
results from adding the sensitivity of the individual
sub-telescopes (\S 5).  Imaging using counts accumulated
over different timescales increases the sensitivity (\S 6).
Finally, we combine these different calculations to
evaluate {\it EXIST}'s overall capabilities to study bursts
(\S 7).

\section{Overview of the {\it EXIST} Mission}

The {\it EXIST} design analyzed here was proposed as the
Black Hole Finder Probe for
NASA's Beyond Einstein program.  In this design,
described in Grindlay (2007),
the mission consists of two arrays
of sub-telescopes.  The 19~High Energy Telescopes (HETs)
will use a Cadmium~Zinc~Telluride (CZT) detector plane,
while the 32~Low Energy Telescopes (LETs) will use a
silicon detector plane.  The spacecraft will be launched
into low Earth orbit ($\sim$500~km) by either a Atlas~V-551
(for an orbital inclination of $i\sim20^\circ$) or a
Delta~IV~4050H ($i\sim5^\circ$) for a 5 (minimum)--10
(goal) year mission. The CsI active shielding for the CZT
detectors will also be instrumented to provide spectral
coverage for gamma-ray bursts at higher energies.  Table~1
provides the detector parameters relevant to this study.
The spacecraft pointing will rock $\pm 15^\circ$
perpendicular to the orbit around the zenith, resulting in
nearly uniform sky coverage and sensitivity.

Both the HETs and LETs will image the gamma-ray sky using
the coded mask technique.  The detector plane `sees' the
sky through a mask with open and closed cells that is a
fixed distance above the detector plane.  Therefore a
source in the FOV casts a shadow with the mask's pattern on
the detector plane. The distribution of sources on the sky
is deconvolved from the counts detected by the position
sensitive detectors.  Sources in the central part of the
FOV, the `fully coded' region, illuminate the full detector
plane, while sources further out in the FOV, the `partially
coded' region, illuminate only a fraction of the detector
plane.  The dimensions of the detector plane and mask, and
the detector-mask distance, determine the FOV, while the
detector-mask distance and the dimensions of the mask cells
and detector pixels fix the angular resolution.

As for {\it Swift} and {\it GLAST}, {\it EXIST} will run
burst detection and localization software onboard (the Fast
Onboard Burst Alert System---FOBAS), and telemeter data to
the ground for further analysis.  In the current design,
FOBAS will run both rate and image triggers on the
datastream from both the HET and LET sub-telescopes. Rate
triggers will search for statistically significant
increases in the count rates from the sub-telescopes.  The
image triggers will form images from the counts from the
individual sub-telescopes, add the images, and search the
resulting sky image for a new, statistically significant
point source.  In the current design, images will be formed
with 3, 18, 108, 648 and 1296~s accumulation times. When a
burst is detected, {\it EXIST} will downlink the burst time
and location (as well as other basic burst parameters)
through the Tracking and Data Relay Satellite System
(TDRSS) within $\sim10$--20~s, as is done for {\it Swift}
and will be done for GLAST.

Data indicating the time, energy and pixel of every HET and
LET count, as well as other standard science and
housekeeping data, will be downlinked approximately every
four hours through a TDRSS Ku band link, as will be done
for GLAST. Ground software will calculate more accurate sky
positions and other parameters (e.g., durations and
spectra) for the bursts detected onboard, and will search
the datastream for bursts that FOBAS did not detect.

The active CsI shields behind the HET detector plane and in the
lower part of the HET collimators will be instrumented to provide 64
channel spectra between $\sim$300~keV to $\sim$10~MeV (see Garson et
al. 2006a). The current plan is that spectra accumulated every 1~s
will be downlinked.  By buffering the counts from the shields, the
count binning will be increased to every 0.1~s for the time period
500~s before to 500~s after the trigger.

\section{Burst Detection Sensitivity}

A burst will be detected by {\it EXIST} when a statistically
significant new source is found in either an HET or LET image of the
sky. The same criterion applies to {\it Swift}'s BAT (a single CZT
coded mask detector) and therefore the sensitivity analysis we use
follows the methodology applied to the BAT in Band (2006).

Formation of the image in which the burst is detected may
be initiated by either a rate or image trigger.  A rate
trigger will search the count rates from the sub-telescopes
for a statistically significant increase.  An image trigger
will search for new sources in images of the sky that will
be formed continuously. The new source in an image trigger
may not be statistically significant, but will indicate
that a burst may be in progress. After either a rate or
image trigger, FOBAS (the burst flight software) will vary
the time and energy ranges over which counts are
accumulated to maximize the signal-to-noise ratio. An image
will then be formed from these counts.  The threshold for
these initial rate or image triggers will be sufficiently
loose to allow many triggers; the absence of a
statistically significant point source in the final image
will weed out the false positives. Note that the final
image in which the point source is most significant may be
formed from the counts in a different energy and time bin
than that of the counts that initially triggered FOBAS.

Regardless of the process leading to the final image, {\it
EXIST}'s burst sensitivity will be the minimum burst flux
that results in a statistically significant point source in
an HET or LET image.  This is the basis of our analysis. In
this section we calculate the sensitivity for a point
source in the center of the FOV of a single sub-telescope,
and in \S 5 we consider how this sensitivity varies across
the sub-telescope arrays' FOV.

Skinner (2007) derived the source detection sensitivity of
a coded mask system when standard assumptions are relaxed:
the fraction of the open mask pixels may differ from 1/2;
part of an open mask pixel may be occulted (e.g., by ribs
around each pixel to support the {\it EXIST} masks); the
detector pixels may not be small relative to the mask
pixels; the source strength may be comparable to the
background; and the closed mask pixels may be partially
transparent (e.g., at high energy).  Here we will consider
the background-dominated case.

Consider an image formed using counts accumulated over $\Delta t$
and $\Delta E$.  The burst spectrum is $N(E,t)$ (ph cm$^{-2}$
s$^{-1}$ keV$^{-1}$).  Let $s$ be the source flux averaged over
$\Delta t$ and integrated over $\Delta E$. Thus
\begin{equation}
s = {1\over{\Delta t}} \int_{\Delta E} dE \int_{\Delta t}
   dt N(E,t) \quad.
\end{equation}
Let the average detector efficiency be
\begin{equation}
\epsilon_0 = { {\int_{\Delta E} dE
   \int_{\Delta t} dt \, \epsilon(E) N(E,t)}\over
   {\int_{\Delta E} dE \int_{\Delta t}
   dt N(E,t)}} \quad ,
\end{equation}
where $\epsilon(E)$ the detector efficiency.  At high
energy photons leak through the closed mask pixels.  Define
an effective detector efficiency for flux through the
closed mask pixels
\begin{equation}
\epsilon_1 = {{ \int_{\Delta E} dE
   \int_{\Delta t} dt \, \epsilon(E) N(E,t) e^{-\tau_m(E)}}\over
   {\int_{\Delta E} dE \int_{\Delta t}
   dt N(E,t)}} \quad ,
\end{equation}
where $\tau_m(E)$ is the optical depth through the closed mask
elements. Note that $\epsilon$ includes absorption by all material
over the entire detector, while $\tau_m(E)$ accounts only for
absorption through the closed mask pixels.

For the HETs Garson et al. (2006b,c) find that the photon
aperture flux from the Cosmic X-ray Background (CXB)
dominates the background below $\sim 100$~keV, while at
higher energy sources such as Earth albedo photons, charged
particles and activation dominate the background. For the
LETs the background results primarily from the CXB. The CXB
contribution can be modelled semi-analytically, while other
background sources require complex Monte Carlo
calculations.  Thus we model the total background count
rate per detector area as
\begin{equation}
B = (\Delta t)^{-1} \int_{\Delta E} dE \int_{\Delta t} dt
   \left[ \epsilon(E) \Phi(E) \Omega_a \left[f_{\rm mask}+
   (1 -f_{\rm mask}) e^{-\tau_m(E)} \right] + b(E) \right]
\end{equation}
where $f_{\rm mask}$ is the fraction of the mask area that
is open, $\Phi(E)$ is the CXB (Gruber 1992), $\Omega_a$ is
the projected solid angle subtended by the detector's
aperture averaged over the detector plane (calculated with
the corrected formulae of Sullivan [1971]), and $b(E)$
models the other sources of background.  The aperture flux
includes the leakage of the CXB through the closed mask
elements at high energy.

The significance of the burst's image in the
background-dominated case is
\begin{equation}
S_I = f_m  s {{(\epsilon_0-\epsilon_1)}\over 2} \sqrt{{A \Delta t}
   \over B} \quad ,
\end{equation}
where $A$ is the detector area, and $f_m$ includes the
factors resulting from the ribs around the mask pixels, the
fraction of the mask pixels that are open, and the finite
detector size.  For open mask pixels where the ribs cover
0.2 of the pixel area and detector-to-mask pixel ratios of
1/2 (HET) and 1 (LET), $f_m=0.737$ and 0.564, respectively.

It is convenient to parameterize the burst flux in terms of
\begin{equation}
F_T = (\Delta t)^{-1} \int_{\Delta E_0} dE \int_{\Delta t}
   dt N(E,t) = s \, {{\int_{\Delta E_0} dE \int_{\Delta t}
   dt N(E,t)}\over {\int_{\Delta E} dE \int_{\Delta t}
   dt N(E,t)}}  \quad.
\end{equation}
where $\Delta E_0$=1--1000~keV.

To understand the effect of transparency through the closed
mask pixels, we define the mask factors
\begin{equation}
g_s = {{1-e^{-\tau_m(E)}}\over 2} \quad,
\end{equation}
(relevant to the factor of $(\epsilon_0-\epsilon_1)/2$),
and
\begin{equation}
g_b = f_{\rm mask}+ (1 -f_{\rm mask}) e^{-\tau_m(E)}
   \quad.
\end{equation}
(relevant to the aperture flux).  A decrease in $g_s$
results in a decrease in the detection significance, while
an increase in $g_b$ indicates an increase in the aperture
flux.

The gamma-ray burst spectrum is modelled using the
four-parameter `Band' function (Band et al. 1993): a low
energy power law with an exponential rolloff ($N(E)\propto
E^\alpha \exp[-E/E_0]$) that merges smoothly with a high
energy power law ($N(E)\propto E^\beta$).  The break
between the two power laws is characterized by
$E_p=(2+\alpha)E_0$, which is the energy of the maximum of
$E^2N(E)\propto \nu f_\nu$ if $\beta<-2$, i.e., $E_p$ is
the photon energy where most of the energy is radiated.  We
use the flux $F_T$ (ph~s$^{-1}$~cm$^{-2}$) integrated over
the 1--1000~keV band to normalize the spectrum (see eq.~6).
Thus the spectrum is characterized by the normalization
$F_T$, the two spectral indices $\alpha$ and $\beta$, and
the energy $E_p$.

Eq.~5 can be inverted to find the threshold value of $F_T$
at the peak of the lightcurve for a given set of the
spectral parameters that determine the shape of the burst
spectrum---the spectral indices $\alpha$ and $\beta$, and
the peak energy $E_p$.  The result is a surface in the four
dimensional space given by these spectral parameters;
bursts with spectra on one side of this surface (with $F_T$
greater than the value on the surface) will be detected,
while bursts on the other side will not. Holding the
spectral indices $\alpha$ and $\beta$ fixed projects this
surface into a sensitivity curve in the $F_T$-$E_p$ plane.
The curve also depends on the accumulation time $\Delta t$
(here $\Delta t=1$). The dependence of the sensitivity on
the accumulation time is discussed below (\S 6).

Note that the threshold value of $F_T$ at a given $E_p$ is
{\it not} the sensitivity of the detector at a photon
energy equal to $E_p$. The power of sensitivity curves in
the $F_T$-$E_p$ plane is that they show the detectability
of a burst with a given set of spectral parameters, and
thus the sensitivity of different detectors can be
compared, regardless of their specific energy response.  A
CZT-based detector that detects photons in the 10--150~keV
band can be compared to a scintillator-based detector that
detects photons in the 50--300~keV band.

\section{Single Sub-Telescope Energy Sensitivity}

\subsection{HET}

A single HET will have a 56~cm $\times$ 56~cm (an area of
3136~cm$^2$) CZT detector plane that is 5~mm thick.  The platinum
and gold cathode pads on the CZT ($\sim 1000$~\AA\ each), the Mylar
thermal blankets (two~$\sim5$~mil blankets) and the Kevlar
micrometeriod shield ($\sim$5~mil) in the current design produce
negligible absorption $>10$~keV.  Figure~1 shows the efficiency of
the CZT detectors as a function of energy.

We calculate the mask factors (eqs.~7 and 8) using the
optical depth through a 5~mm thick, 107.9~cm $\times$
107.9~cm plate of tungsten. Because of the supports
necessary for the closed mask pixels, we assume an open
fraction of $f_{\rm mask}$ = 0.4.  Figure~2 shows the
resulting mask factors as a function of energy for both the
nominal 5~mm thickness (solid curve) and for a mask with
half this thickness (dashed curve).

The linear dimensions of the detector and mask pixels will be
0.125~cm and 0.25~cm, respectively.  For a detector to mask pixel
dimension ratio of 1:2 the factor compensating for the finite size
of the detector pixels is $f_m=0.737$ (see eq.~5).

The background is modelled (eq.~4) as the sum of the CXB aperture
flux and the continuum background from other sources (see Garson et
al. 2006b,c for details). Figure~3 shows the resulting background.

We assume that counts are accumulated over two energy bands
$\Delta E$ = 10--600 and 40--600~keV.  The 10--600~keV band
is sensitive to soft bursts, where the large number of low
energy burst counts compensates for the large aperture
flux, while the 40--600~keV band is particularly sensitive
to hard bursts where there are sufficient burst photons
above 40~keV. The required threshold significance is
assumed to be $S_I = 7$ (the same as the BAT's threshold).

Figure~4 shows the sensitivity curve for a single HET sub-telescope
for three sets of spectral indices and for $\Delta t$=1~s.

The survey will localize sources to better than 56~arcsec
(Grindlay 2007).  The survey's threshold will be
5$\sigma$ whereas the burst threshold will be 7$\sigma$,
and typically localization is proportional to $\sim
(\sigma-1)^{-1}$.  Thus we estimate that the HET's
localizations will be better than 40~arcsec.

\subsection{LET}

As currently designed, a single LET will have a 20~cm $\times$ 20~cm
(an area of 400~cm$^2$), 1~mm thick Si detector plane with 0.02~cm
pixels.  Figure~1 also shows the LET efficiency.  The mask will be
72~cm above the Si detector plane, with collimators extending from
the detector plane to the mask.  The 20~cm $\times$ 20~cm mask will
have a thickness of 0.05~mm and 0.02~cm pixels.  Again, because of
the need to support the closed mask pixels, we assume an open
fraction of $f_{\rm mask}$=0.4.  Over the LET's energy range (3 to
30~keV) the closed mask pixels will be optically thick. With a mask
to detector pixel ratio of 1:1, the mask factor in eq.~5 is
$f_{m}$=0.564.

Although we include an internal background of $b(E)=10^{-5}$ cts
cm$^{-2}$ keV$^{-1}$ s$^{-1}$ in our calculation, the background is
almost entirely the result of the CXB aperture flux; see Figure~3.

We assume a threshold image significance of $S_I=7$ will be required
over a single trigger energy band of $\Delta E$= 3--30~keV. Figure~4
shows the resulting sensitivity.  Note that the LETs are less
sensitive than the HETs.  The slopes of the HET and LET sensitivity
curves are consistent with the energy dependence of the LET and HET
detectors.

The EXIST survey's localizations should be better than 11
arcsec (Grindlay 2007), and thus accounting for the
difference in survey and burst thresholds (5$\sigma$ vs.
7$\sigma$), the LET's burst localizations should be better
than 8~arcsec.

\section{Off-Axis and Multi-Detector Sensitivity}

The arrays of HET and LET sub-telescopes will each cover a
very large total FOV. Any point in these total FOVs will be
in the fully- or partially-coded FOVs of a number of
sub-telescopes.  The resulting multi-detector sensitivity
across the arrays' FOVs will depend on how the images from
the different detectors will be added together; this
merging will depend on the exigencies of the available
computational power and the required data latency.
Specifically, burst detection and localization on-board the
{\it EXIST} spacecraft by radiation-hardened processors
will probably be less sensitive than on the ground, where
farms of high-speed processors will be available. In
addition, localization on-board must be rapid so that
telescopes on the ground can begin following the burst
afterglow.

The calculations above provide the on-axis sensitivity for
single HET or LET sub-telescopes.  Let $R$ be the ratio of
the actual sensitivity at a given point in the FOV to this
single sub-telescope on-axis sensitivity, where sensitivity
is proportional to $S_I$ (see eq.~5) or to the inverse of
the threshold peak flux $F_T$.  Thus larger $R$ means a
greater significance for a given peak flux or a smaller
threshold peak flux for a given significance.

The source flux falling on the detector plane is only a
fraction $f_c \cos\theta$ of the flux it would have
on-axis, where $f_c$ is the `coding fraction' which
accounts for the partial shadowing of the detector plane by
the collimators (the detector sides) and $\theta$ is the
inclination angle (the angle between the source direction
and the detector normal). The non-source flux that
contributes to the background around the source is
proportional to the `coding fraction' $f_c$.  In coded mask
imaging only the counts in the region of the detector plane
that is not shadowed for a given source contribute to the
image around the source. The source flux that impinges on
this region is foreshortened by the inclination angle (the
`$\cos\theta$' effect), but the background in this region
does not depend on the source's direction.  For a single
sub-telescope the ratio of the off-axis to on-axis
significance is therefore $R=f_c^{1/2} \cos\theta$.

The methodology by which the data from multiple detectors
will be combined is currently being studied.  First, the
images can be added.  Then the source flux is proportional
to $\sum f_{c,i} \cos\theta_i$, the background to $\sum
f_{c,i}$, and thus $R_I = [\sum f_{c,i} \cos\theta_i]
/\sqrt{\sum f_{c,i}}$.  Alternatively, forming images for
each sub-telescope and adding the significances for the
common point sources in quadrature gives $R_Q=\sqrt{\sum
f_{c,i} \cos^2\theta_i}$.  In practice for the HET and LET
arrays the sensitivity over the FOV for these two methods
differ very little, and we use $R_I$.

To calculate the sensitivity over the FOV, we work, and
plot results, in a coordinate system that is a projection
of the spherical sky directly onto a plane perpendicular to
the zenith, i.e., if a point on the sky has the coordinates
$x$,$y$,$z$ (where $\sqrt{x^2+y^2+z^2}=1$), then we work in
the $x$-$y$ plane. In this coordinate system, $z$ is along
the spacecraft's zenith, $x$ is along the direction of
orbital motion, and the spacecraft nods (rocks) in the $y$
direction.  We calculate $R_I$ at different points on this
grid.

Figure~5 shows the burst sensitivity over the sky for the HET and
LET arrays; the maxima are just under twice the sensitivities (i.e.,
more sensitive than) of single HET and LET sub-telescopes. Figure~6
shows the amount of solid angle at a given sensitivity for both
arrays.  Thus different points in the overall FOV will have
different sensitivity thresholds, which must be considered when
analyzing the cumulative intensity distribution.  Figure~7 shows the
low end of the cumulative intensity distribution resulting from
variations in the threshold over the FOV; other effects that smooth
the threshold are ignored, and therefore the effect demonstrated by
this figure applies to any burst intensity distribution. Note that
the sensitivity of {\it Swift}'s BAT also varies over the FOV,
affecting the shape of the cumulative fluence or peak flux
distributions.

\section{Dependence on Accumulation Time $\Delta t$}

The HET and LET sensitivity curves presented in Figure~4 assumed
$\Delta t$=1~s, i.e., that the bursts were detected in images formed
over 1~s.  However, modern burst detectors (e.g., {\it Swift}'s BAT,
the GLAST Burst Monitor and {\it EXIST}) usually use a number of
different accumulation times.  For an imaging detector the relevant
$\Delta t$ is the accumulation time for the final image. An
accumulation time comparable to the burst duration will usually
maximize the source significance. A longer accumulation time will
dilute the signal with background, reducing the signal-to-noise
ratio, and therefore the significance of the detection.  On the
other hand, a shorter accumulation time will often exclude signal
that could have increased the significance of the burst detection.

Quantitative analysis of the dependence of burst sensitivity on the
accumulation time is difficult because of the large range of burst
durations and the great diversity of burst lightcurves.  Some bursts
consist of contiguous, overlapping pulses while others have widely
separated pulses.  Band (2002) ran a software rate trigger with a
wide range of $\Delta t$ values on the lightcurves of 100 bright
BATSE bursts, and determined that using a range of $\Delta t$ values
would increase the burst detection rate by $\sim$25\% over the rate
for $\Delta t$=1~s.  Band (2006) explained the larger fraction of
long duration bursts relative to short duration bursts in the {\it
Swift} data set compared to BATSE's as resulting in part from {\it
Swift}'s long accumulation times.

As a demonstration of the increase in sensitivity afforded by using
a variety of accumulation times, consider a burst lightcurve with an
exponential shape, $N(t)=N_0 \exp[-t/T]$; the traditional duration
of 90\% of the emission is $T_{90}=T\,\ln 10$. In this example the
accumulation time is assumed to begin at $t=0$.  Let $F_T(\Delta t)$
be the threshold peak flux averaged over 1~s (this is the quantity
plotted in Figure~4 for $\Delta t$=1~s) for a given $\Delta t$. Then
the ratio of threshold peak fluxes for two different accumulation
times $\Delta t_0$ and $\Delta t_1$ is
\begin{equation}
{{F_T(\Delta t_1)}\over{F_T(\Delta t_0)}} =
   \sqrt{{\Delta t_1}\over{\Delta t_0}}
   {{1-\exp[-\Delta t_0/T]}\over{1-\exp[-\Delta t_1/T]}} \quad .
\end{equation}
If a detector uses a set of $\Delta t$ values, then the
smallest value of $F_T(\Delta t)$ should be used for any
given value of $T$.  We assume we are in the
background-dominated case (eq.~5); the detectability of
very short bursts might be limited by a paucity of source
counts.

Figure~8 shows this ratio for $\Delta t_0$=1~s and different sets of
$\Delta t_1$.  Thus this figure shows how the sensitivity of a
mission such as {\it EXIST} to short and long duration bursts is
increased by using a variety of accumulation times.  The dashed
curve assumes $\Delta t_1$=1~s, and thus the ratio is equal to 1.
Currently {\it EXIST}'s planned imaging trigger (which is {\it not}
the final imaging step in {\it EXIST}'s burst detection process)
will use $\Delta t_1$=\{3, 18, 108, 648 and 1296\}~s; this is shown
by the solid curve.  Finally, $\Delta t_1$ may be varied to maximize
the signal-to-noise ratio, minimizing $F_T(\Delta t_1)$ to the
smallest possible value; this is shown by the dot-dashed curve.

If burst lightcurves could be described by the exponential shape of
this example (and bursts did not undergo spectral evolution, which
makes the duration energy-dependent), then the HET or LET threshold
peak flux of a burst of a given peak energy $E_p$ and duration
$T_{90}$ would be the product of $F_T$ from Figure~4 and the ratio
from Figure~8.

We emphasize that this is a highly idealized example meant to
demonstrate how the variable accumulation times of {\it EXIST}'s burst detection system will increase the sensitivity to long and short duration
bursts.  This is particularly relevant to high redshift bursts whose
durations will be time-dilated.

\section{Discussion}

From the preceding analysis, we can draw several
conclusions on {\it EXIST}'s impact on the study of
gamma-ray bursts.

First we estimate the {\it EXIST} burst detection rate. The
BATSE observations provide the cumulative burst rate as a
function of the peak flux value $\psi_B$ averaged over
$\Delta t=1$~s in the $\Delta E=$50--300~keV band (Band
2002):
\begin{equation}
N_B \sim 550 \left[ {{\psi_B}\over \hbox{0.3 ph cm$^{-2}$
   s$^{-1}$}}
   \right]^{-0.8} \hbox{ bursts yr$^{-1}$ sky$^{-1}$ } \quad .
\end{equation}
The HET threshold sensitivity for a single sub-telescope on-axis is
$\psi_B\sim 0.12$~ph~cm$^{-2}$~s$^{-1}$ for $E_p > 100$~keV. Using
the BATSE rate in eq.~10 and integrating over the solid angle
distribution in Figure~6 gives a burst detection rate for the HET of
$\sim 400$ bursts per year. Note that this rate is over the
BATSE-specific values of $\Delta E$ and $\Delta t$, and {\it EXIST}
will use at least two different values of $\Delta E$ (see \S 4.1)
and a variety of $\Delta t$ values (see \S 6). Consequently this
rate should be increased by approximately 50\% to account for the
soft, faint, long duration bursts to which BATSE was less sensitive
than {\it EXIST}'s HET will be; we therefore expect the HET array to
detect $\sim 600$~bursts per year.

The value of $\psi_B$ for an LET varies more with the burst spectral
parameters than for an HET, and therefore estimates of the LET burst
detection rate based on the BATSE rate are much more uncertain.  For
a single LET $\psi_B\sim 0.3$~ph~cm$^{-2}$~s$^{-1}$ on axis at
$E_p=100$~keV, which gives a burst detection rate of $\sim$180
bursts per year using eq.~10 and the LET distribution in Figure~6.
This rate should be increased by a factor of 2 to account for the
different energy band $\Delta E$ and accumulation times $\Delta t$.
We use a larger adjustment factor for the LETs than for the HETs
because the LETs' energy band will overlap less with BATSE's than
the HETs'.  We therefore expect the LET array to detect
$\sim350$~bursts per year.

Next we simulate the spectra that the {\it EXIST} suite of detectors
will observe.  Figure~9 shows a count spectrum (counts s$^{-1}$
keV$^{-1}$) for a moderately strong burst as it might be observed by
the LETs (lefthand set of curves), HETs (middle set) and the CsI
active shields for the HETs (righthand set; based on Garson et al.
2006a). The solid curves show the signal count rate, while the
dashed curves provide the estimated background.  Thus {\it EXIST}
will facilitate spectral-temporal studies.

Particularly important to physical burst emission models is
determining $E_p$, which is typically of order 250~keV
(Kaneko et al. 2006).  In addition, correlations of $E_p$
with other burst properties, such as the `isotropic' energy
(the Amati relation---Amati 2006) or total energy (the
Ghirlanda relation---Ghirlanda, Ghisellini \& Lazzati 2004),
have been proposed. `Pseudo-redshifts' calculated from the
observables related to the burst-frame parameters in these
relations can be used in burst studies when spectroscopic
redshifts are not available, and can guide ground observers
in allocating telescope time to observing potential high
redshift bursts.  The recently proposed Firmani relation
(Firmani et al. 2006) correlates $E_p$, the peak
luminosity, and a measure of the burst duration, all of
which are related to observables in the gamma ray band.
Thus pseudo-redshifts will be estimated using the Firmani
relation based on {\it EXIST} data alone, independent of
observations by other facilities.

With well determined broadband spectra down to 3~keV, {\it
EXIST} will be capable of determining whether the Band
function (Band et al. 1993) suffices to describe burst
spectra.  For example, Preece et al. (1996) found evidence
in the BATSE data for the presence of additional emission
below 10~keV.

By scaling from the EXIST survey's source localization
(Grindlay 2007), we find that bursts should be
localized at threshold by the HETs and LETs to better than
40~arcsec and 8~arcsec, respectively; this localization
should scale as $\sim(\sigma-1)^{-1}$.  Because the HETs
are more sensitive to the LETs, the HET localization is
relevant to the faintest bursts EXIST will detect.

{\it EXIST}'s burst capabilities calculated above will constitute a
major leap beyond current detectors, and should increase the number
of high redshift bursts detected.  On average, high redshift bursts
should be fainter, softer and longer than low redshift bursts
(although the broad burst luminosity function and great variety in
burst lightcurves and spectra obscure this trend).  Figure~10
compares the detector sensitivities of the HET (solid curve) and LET
(dashed curve) arrays to the BAT on {\it Swift} (dot-dashed curve)
and BATSE's Large Area Detector (LAD---dot-dot-dashed curve).  As
discussed above, the sensitivity is the threshold peak flux $F_T$
integrated over the 1--1000~keV band as a function of the spectrum's
$E_p$; $\alpha=-1$ and $\beta=-2$ are assumed. In addition, the
figure shows families of identical bursts at different redshifts
(the curves with the points marked by `+'). Each family is defined
by the value of $E_p$ in the burst frame; here again $\alpha=-1$ and
$\beta=-2$ are assumed. In each family the burst would be observed
to have $F_T$=7.5 ph cm$^{-2}$ s$^{-1}$ if it were at $z=1$. The
points marked by `+' are spaced every $\Delta z=1/2$; thus the
uppermost points are at $z=1$ and the lowermost points are at
$z=10$. The pulses in burst lightcurves become narrower (shorter) at
higher energy, an effect that is generally proportional to $E^{0.4}$
(Fenimore et al. 1995). Since the observed lightcurve originated in
a higher energy band, pulses should become narrower with redshift,
reducing the peak flux when integrated over a fixed accumulation
time; the plotted families include this effect.  Finally, in \S 6 we
showed that forming images on long timescales increases the
sensitivity to long duration bursts, as might result from
cosmological time dilation.

\section{Summary}

We presented our method for analyzing the gamma-ray burst sensitivity of
{\it EXIST}, and applied it to the design for the Beyond Einstein
program; this methodology will be used to guide and evaluate the
evolving mission design.
With two arrays of coded mask detectors covering the
3--30~keV (Si) and 10--600~keV (CZT) bands and non-imaging
high energy CsI detectors (0.2--10~MeV), {\it EXIST} will
be a significant gamma-ray burst observatory.  {\it EXIST}
will detect and localize $\sim 2$ bursts per day, observing
their spectra from 3~keV to over 10~MeV.  For bursts with
comparable spectra and lightcurves {\it EXIST} will be
approximately four times more sensitive than {\it Swift}'s
BAT with a much larger FOV.  With these capabilities, {\it
EXIST} will accumulate a large sample of bursts with well
determined properties such as $E_p$ and redshift,
facilitating physical modelling and population studies, and
realizing the potential of gamma-ray bursts as cosmological
probes.

{}

\clearpage

\begin{figure}
\plotone{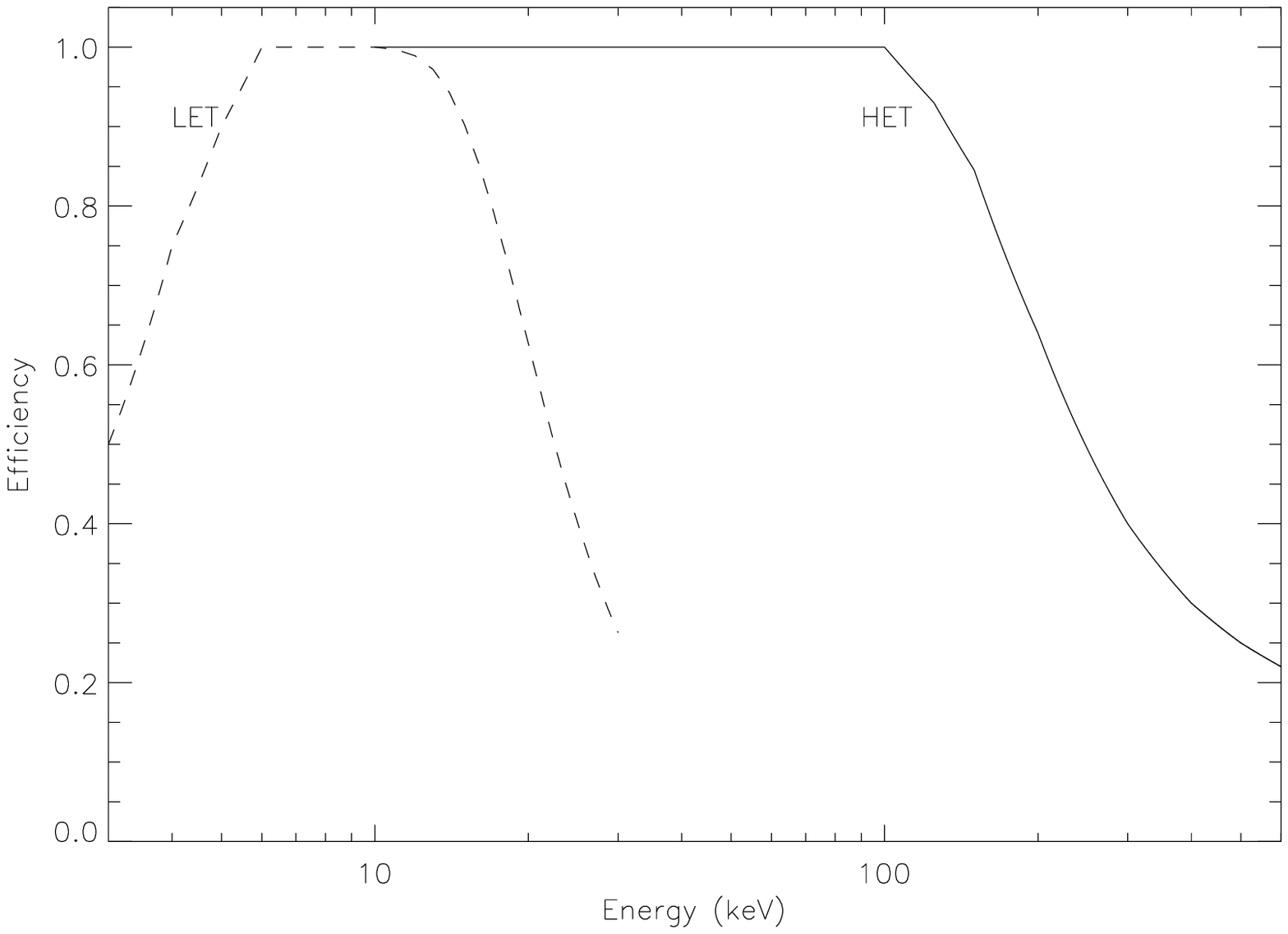} \caption{Efficiency of the 5~mm thick CZT detectors
(solid curve) and the LET detectors (dashed curve) as a function of
energy.}
\end{figure}

\clearpage

\begin{figure}
\plotone{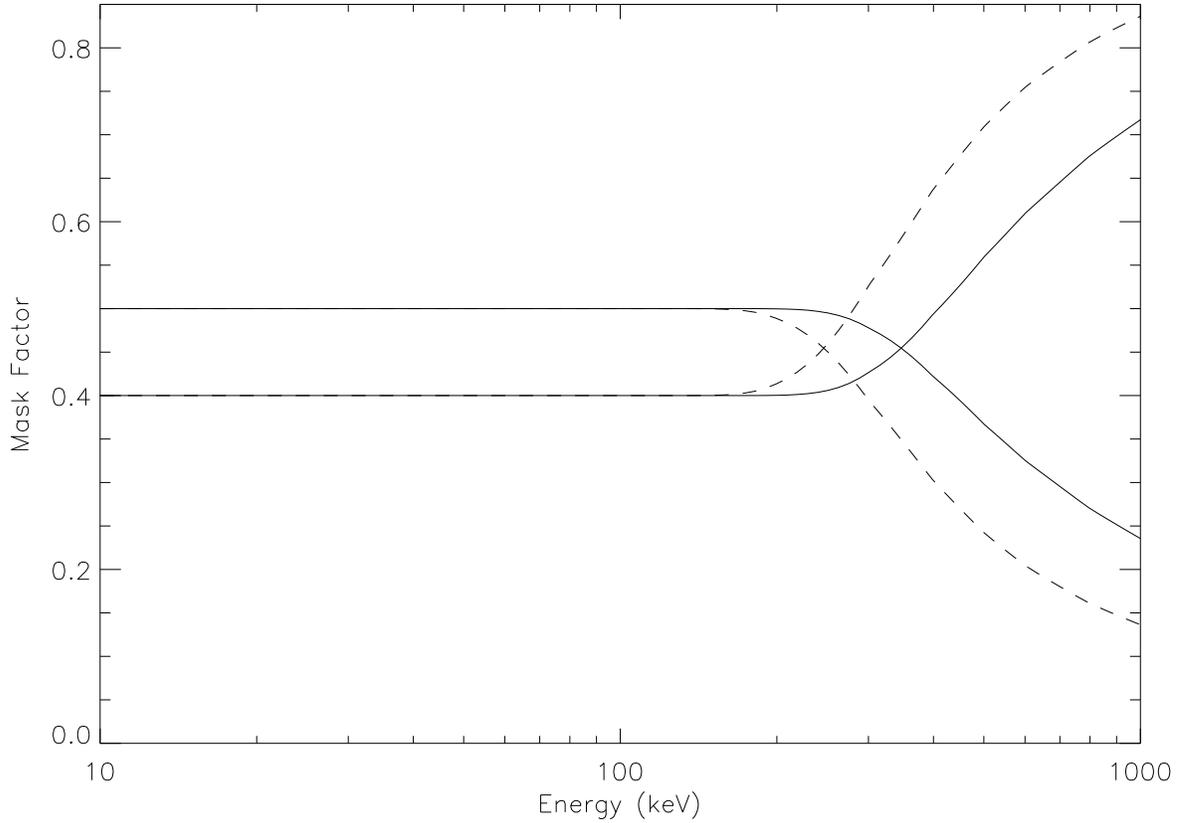} \caption{HET mask factors as a function of
energy for a tungsten mask that is 5~mm thick (solid
curves) and 2.5~mm thick (dashed curves). The set of curves
beginning at 0.4 at low energy is the mask factor $g_{b}$
(eq.~8), the fraction of the incident burst flux that
reaches the detector plane, whether through an open or
closed mask pixel.  The set beginning at 0.5 at low energy
is the mask factor $g_{s}$ (eq.~7), the fraction of burst
flux that will be attributed to the burst when an image is
formed.}
\end{figure}

\clearpage

\begin{figure}
\plotone{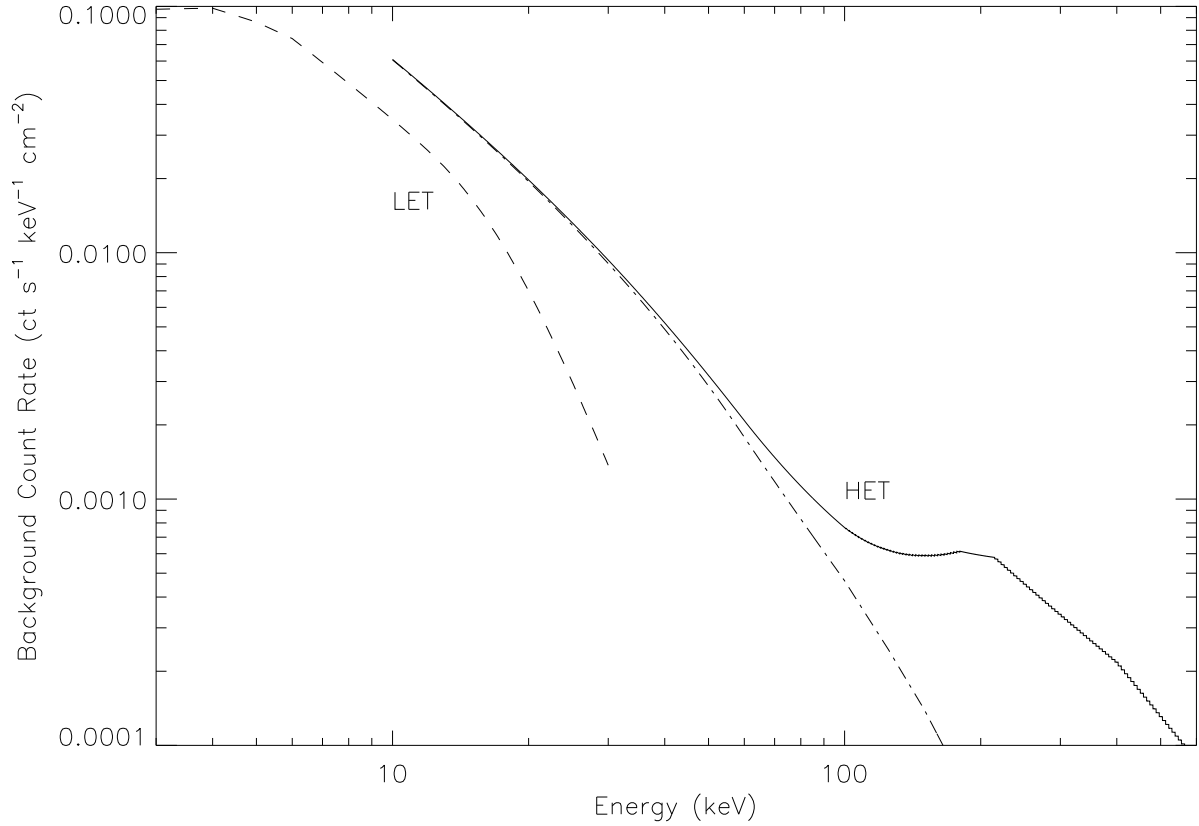} \caption{The background spectrum for the HETs
(solid and dashed-dot curves) and LETs (dashed curve). The cosmic
X-ray background (CXB) aperture flux (dashed curve for LET and
dot-dashed curve for HET) is based on the parameterization of Gruber
(1992) while the total HET background (solid curve) adds other
background components from Monte Carlo simulations (Garson et al.
2006b,c).}
\end{figure}

\clearpage

\begin{figure}
\plotone{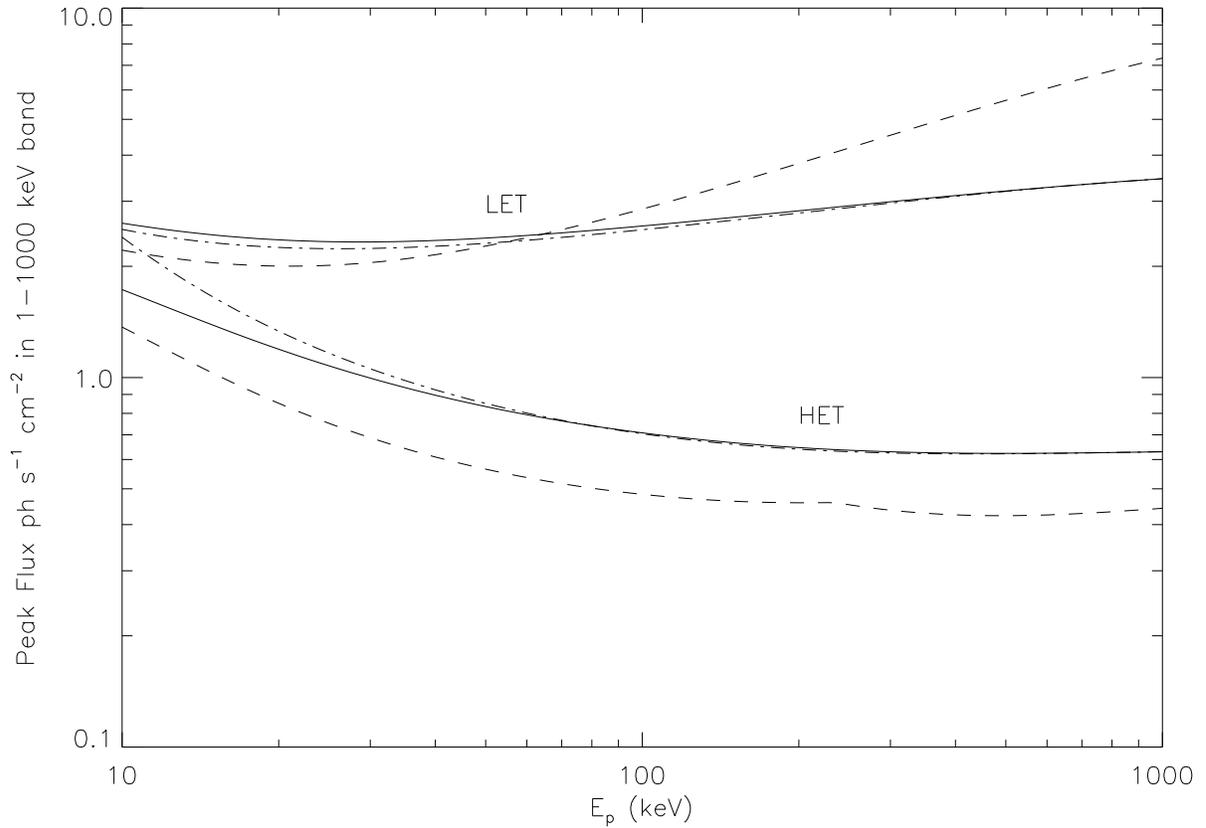} \caption{Maximum HET (lower set of curves) and LET
(upper set) detection sensitivities for $\Delta t=1$~s, the
threshold peak flux over 1--1000~keV as a function of the spectrum's
$E_p$. Solid line---$\alpha = -1$, $\beta = -2$; dashed
line---$\alpha = -0.5$, $\beta = -2$; dot-dashed line---$\alpha =
-1$, $\beta = -3$. Note that this figure shows the sensitivity for
detecting a burst with a spectrum characterized by $E_p$, and not
for detecting a photon with an energy equal to $E_p$. }
\end{figure}

\clearpage

\begin{figure}
\plottwo{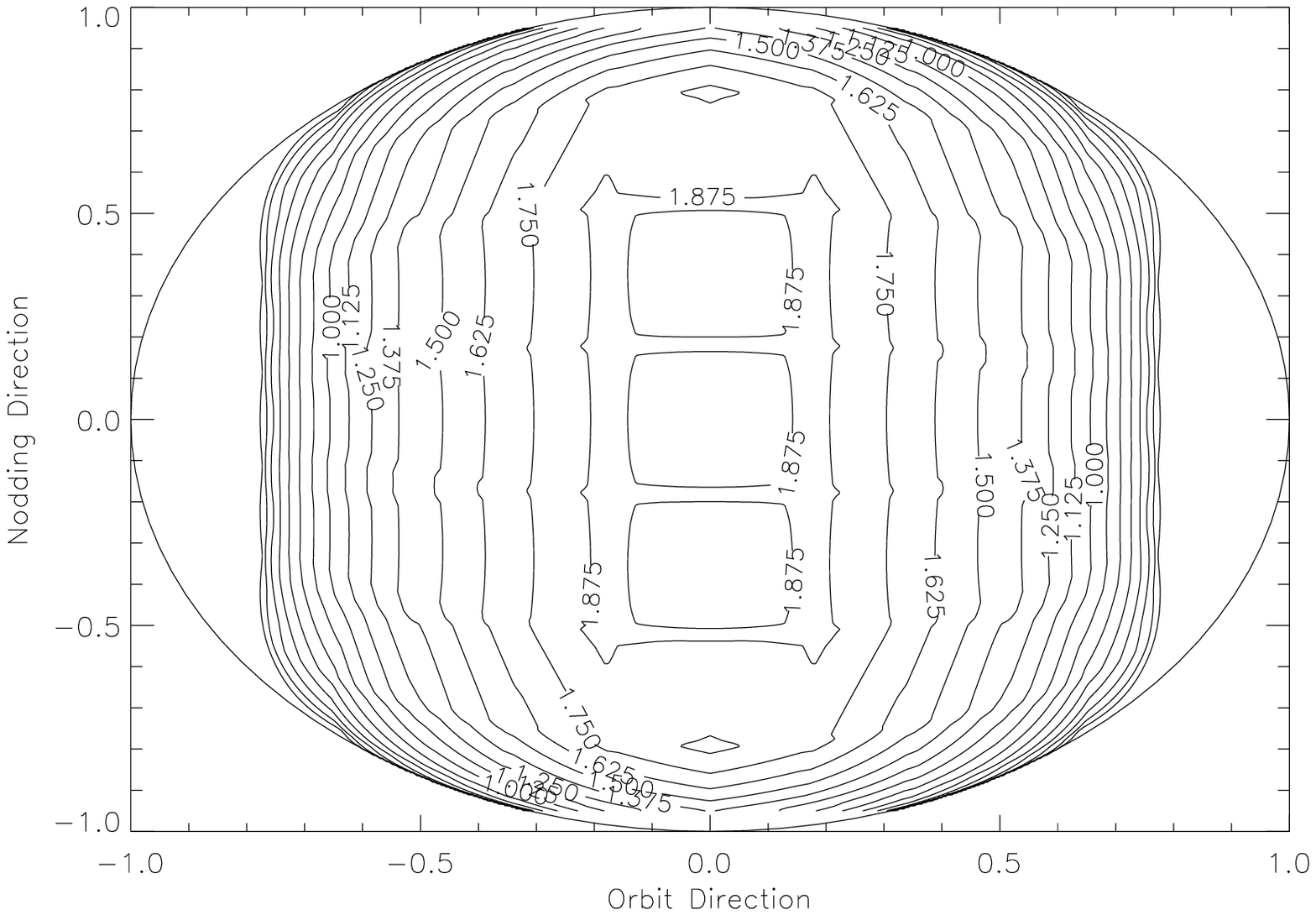}{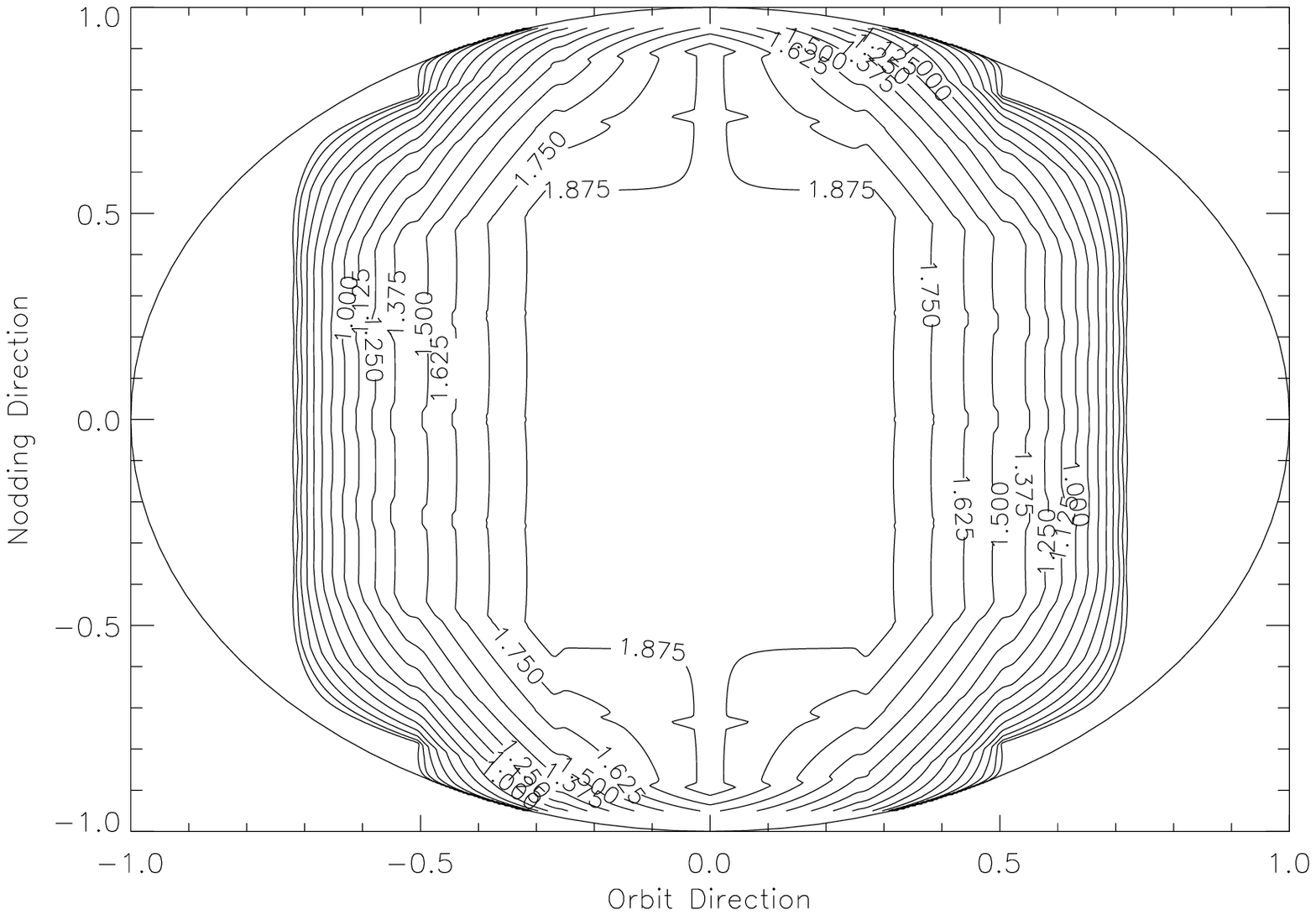} \caption{Burst detection sensitivity over
the sky for the HET (left) and LET (right) arrays, in units of the
sensitivity of a single sub-telescope on-axis.  Greater sensitivity
results in a smaller threshold flux. }
\end{figure}

\clearpage

\begin{figure}
\plottwo{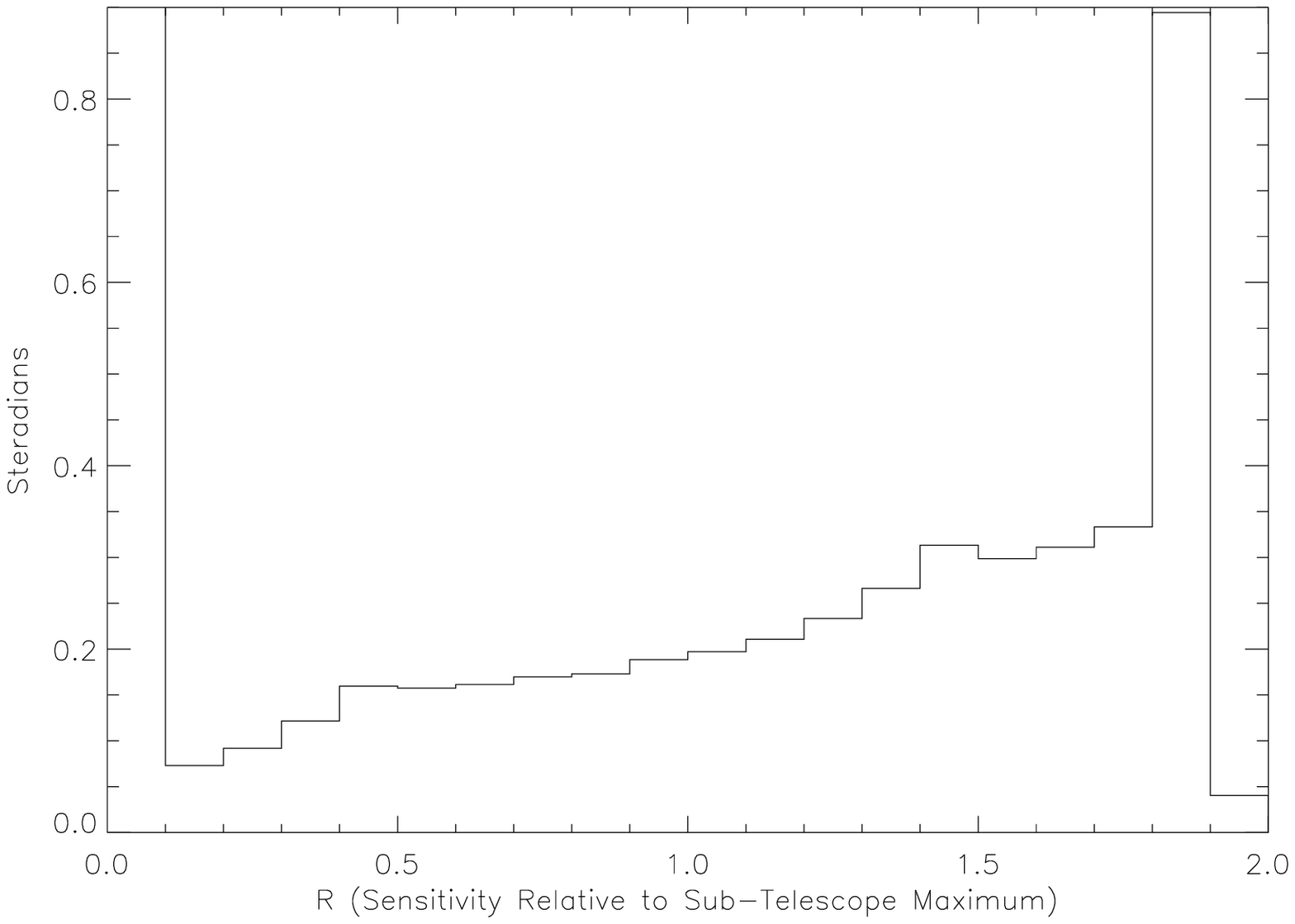}{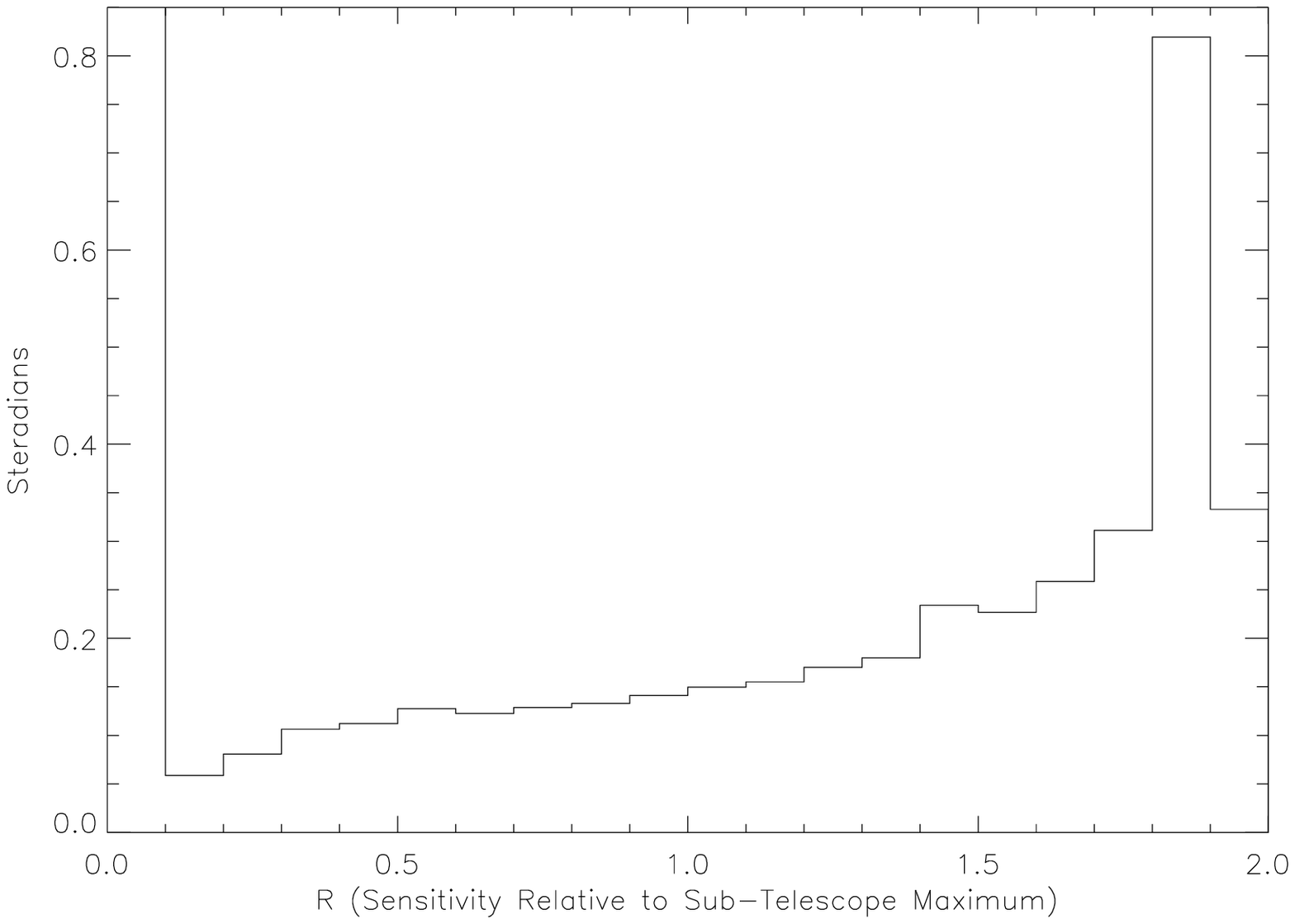} \caption{Solid angle at a given burst
sensitivity for the HET (left) and LET (right) arrays, in units of a
single sub-telescope. Greater sensitivity (larger value of $R$)
results in a smaller threshold flux (smaller $F_T$).}
\end{figure}

\clearpage

\begin{figure}
\plotone{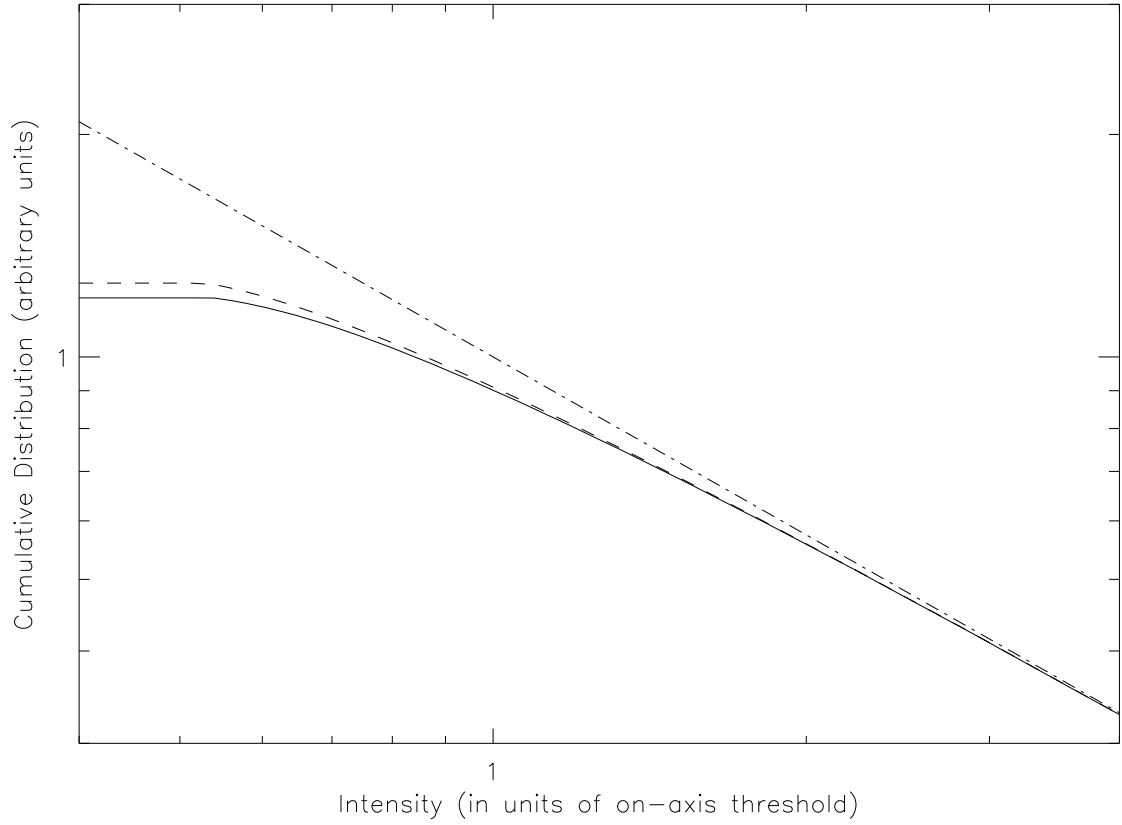} \caption{Effect of the variable detection threshold
across the FOV on the cumulative intensity distribution for the HET
(solid curve) and LET (dashed) arrays.  The assumed actual
distribution (dot-dashed curve) is a power law with an index of
-0.8. The intensity is given in units of the threshold value for a
single sub-telescope. }
\end{figure}

\clearpage

\begin{figure}
\plotone{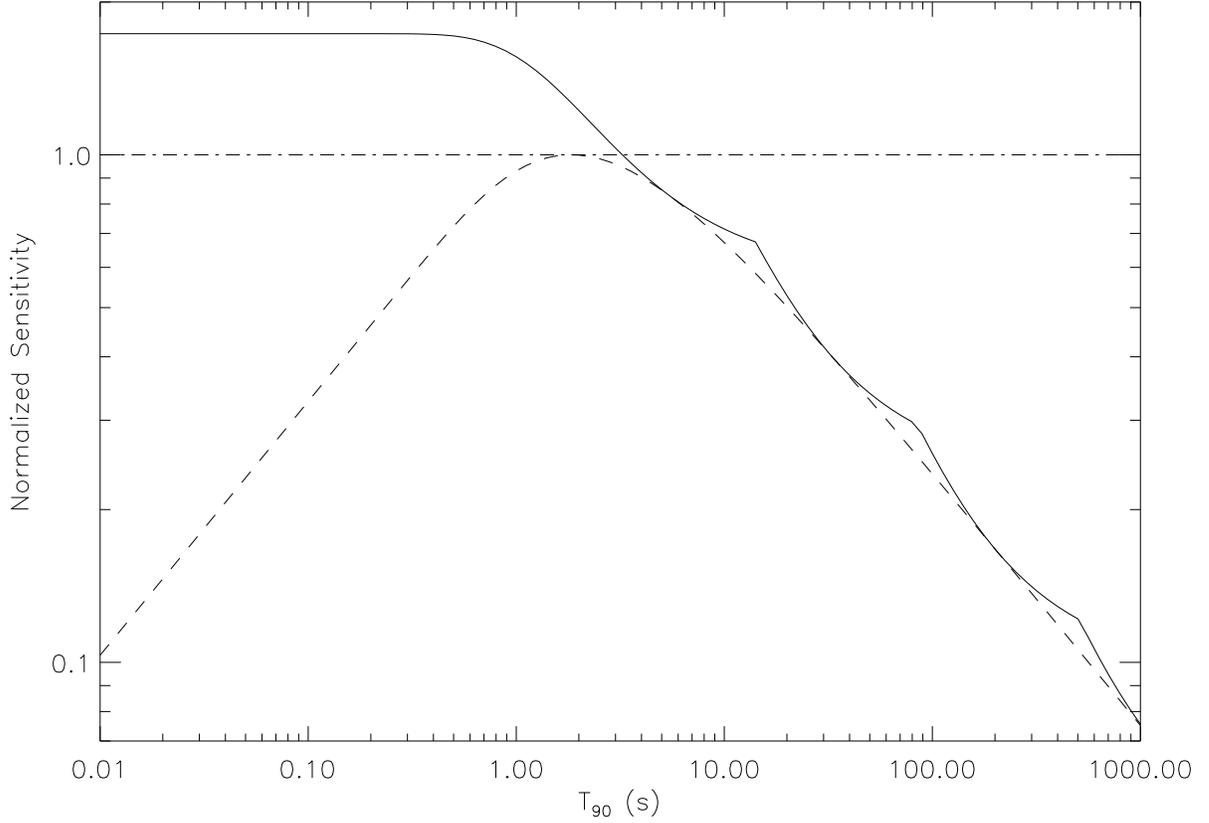} \caption{Normalized burst sensitivity as a function
of burst duration for different accumulation times $\Delta t$.  The
normalized burst sensitivity is the ratio of the peak flux averaged
over 1~s for $\Delta t$=1~s to the peak flux for different $\Delta
t$ values; a smaller value means the detector is more sensitive. The
lightcurve is assumed to be an exponential in time. When a detector
system employs more than one $\Delta t$ then the minimum normalized
sensitivity (resulting from the smallest peak flux) is used.  The
dot-dashed curve assumes $\Delta t$=1, and therefore is 1 for all
durations. The solid curve shows the accumulation times currently
planned for {\it EXIST}'s image triggers: $\Delta t$=3, 18, 108, 648
and 1296~s.  The dashed curve assumes all possible values of $\Delta
t$ and thus shows the smallest possible value of the normalized
sensitivity.}
\end{figure}

\clearpage

\begin{figure}
\plotone{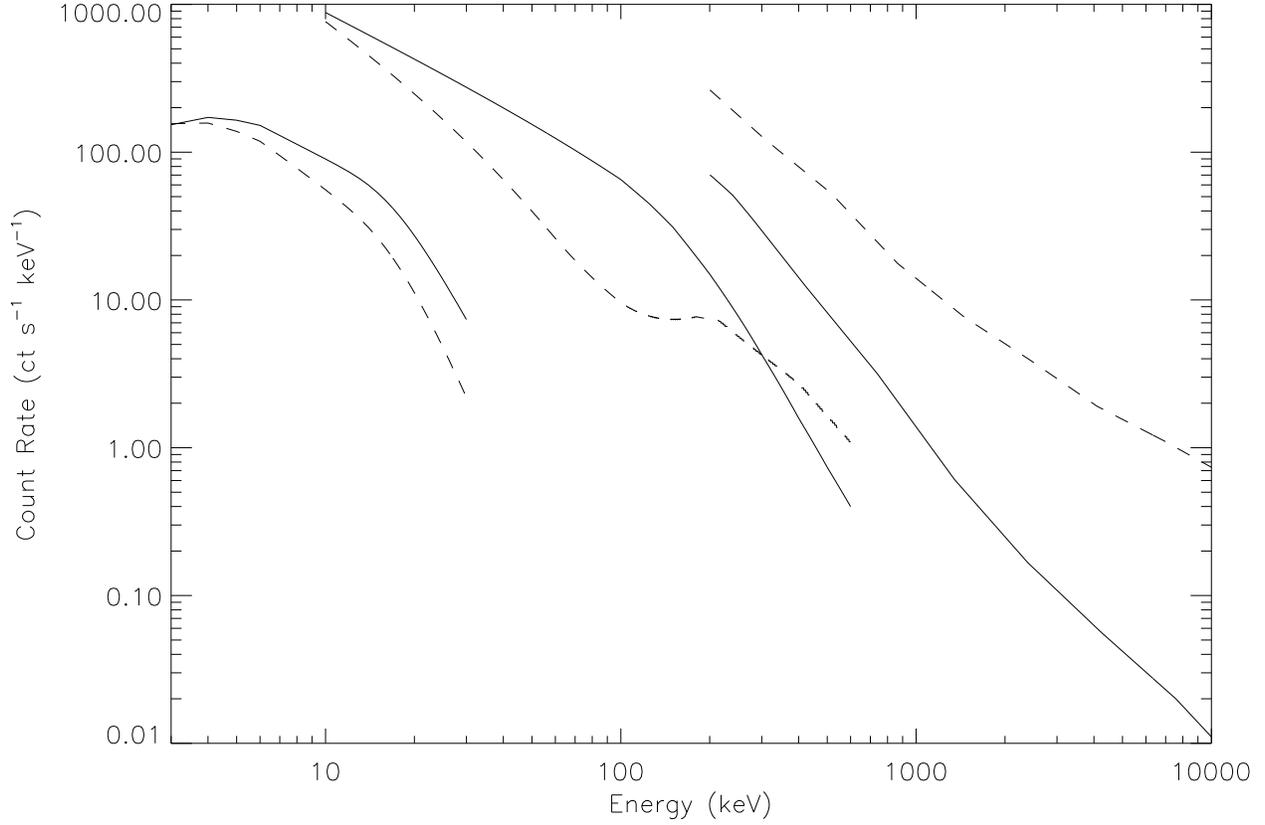} \caption{Source count (solid curves) and background
(dashed curves) spectra for the LETs (left hand set of curves), HETs
(middle set) and CsI shields (right hand set).  The burst spectrum
has $\alpha=-1$, $\beta=-2$, $E_p=300$~keV and
$F_T=7.5$~ph~cm$^{-2}$~s$^{-1}$.  Based on the sub-telescopes' FOVs
in the current design, we assume spectra can be formed from the
equivalent of four HET and LET sub-telescopes, and the shields of
nine HET sub-telescopes.}
\end{figure}

\clearpage

\begin{figure}\plotone{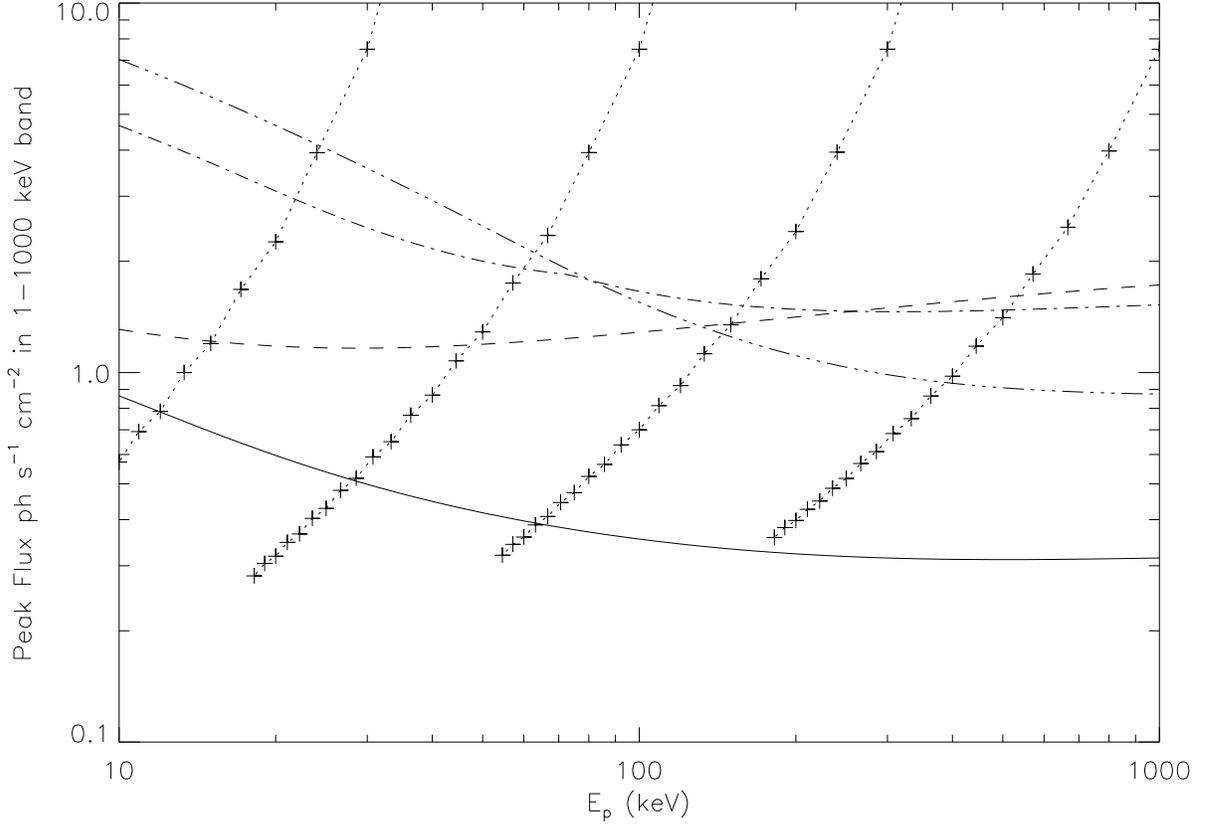} \caption{Maximum detector
sensitivity for HET (solid curve), LET (dashed curve), {\it Swift}'s
BAT (dot-dashed curve) and BATSE's LAD (dot-dot-dashed curve)
assuming $\Delta t=1$~s, $\alpha = -1$, and $\beta = -2$.  The HET
and LET sensitivities assume burst detection by multiple
sub-telescopes. Also shown are tracks for identical bursts at
different redshifts. The bursts have different $E_p$ and $F_T$=7.5
ph cm$^{-2}$ s$^{-1}$ when at $z=1$. The points on the track are
spaced by $\Delta z=1/2$; the faintest bursts on each track are at
$z=10$. Burst pulses are assumed to narrow by $E^{-0.4}$. The
assumed cosmology is $\Omega_m=0.3$ and $\Omega_\Lambda=0.7$ ($H_0$
is irrelevant to this calculation). }
\end{figure}

\clearpage

\begin{deluxetable}{l c c}
\tablecolumns{3}

\tablecaption{\label{Table1}Parameters of the {\it EXIST} Detectors}

\tablehead{
\colhead{Parameter} & \colhead{High Energy Telescope (HET)} &
\colhead{Low Energy Telescope (LET)} }
\startdata
Number & 19 & 32 \\
Detector Material & CZT & Si \\
Detector Thickness & 0.5 cm & 0.1 cm \\
Detector Plane Dimensions & 56~cm $\times$ 56~cm & 20~cm $\times$ 20~cm\\
Detector Pixels & 0.125 cm & 0.02 cm \\
Mask Material & Tungsten & Tungsten \\
Mask Thickness & 0.5 cm & 0.05 cm \\
Mask Dimensions & 107.9~cm $\times$ 107.9~cm & 40~cm $\times$ 40~cm\\
Mask Pixels & 0.25 cm & 0.02 cm\\
Detector-Mask Distance & 140 cm & 72 cm \\
Angular Resolution (FWHM) & 6.86$^\prime$ & 1.35$^\prime$ \\
Localization (7$\sigma$, 90\% conf.) & $<40^{\prime\prime}$
&
   $<8^{\prime\prime}$ \\
Fully Coded FOV & $21^\circ \times 21^\circ$ & $16^\circ
   \times 16^\circ$ \\
Trigger Band $\Delta E$ & 10--600 keV& 3--30 keV \\
 & 50--600 keV & \\
$f_{\rm mask}$ & 0.4 & 0.4 \\
$f_m$ & 0.737 & 0.564 \\
\enddata
%

\end{deluxetable}

\end{document}